\documentclass[conf]{new-aiaa}
\usepackage[utf8]{inputenc}

\usepackage{graphicx}
\usepackage{amsmath}
\usepackage[version=4]{mhchem}
\usepackage{siunitx}
\usepackage{longtable,tabularx}
\usepackage{booktabs}
\usepackage{multirow}
\usepackage[table,xcdraw]{xcolor}
\usepackage{gensymb}
\title{Artificial Hair Sensor Placement Optimization on Airfoils for Angle of Attack Prediction}

\author{Alex C. Hollenbeck\footnote{PhD Candidate, Department of Aeronautics and Astronautics, alex.hollenbeck.2@us.af.mil, AIAA Member, \textit{corresponding author}}, Ramana Grandhi\footnote{Professor, Department of Aeronautics and Astronautics, ramana.grandhi@afit.edu, and AIAA Fellow}, and John H. Hansen\footnote{Assistant Professor, Department of Aeronautics and Astronautics, john.hansen@afit.edu, AIAA Member}}
\affil{Air Force Institute of Technology, Wright-Patterson Air Force Base, Ohio, 45433}
\author{Alexander M. Pankonien\footnote{Research Engineer, Aerospace Systems Directorate, alexander.pankonien@us.af.mil, AIAA Member}}
\affil{Air Force Research Laboratory, Wright-Patterson Air Force Base, Ohio, 45433}

\begin{document}

\maketitle

\begin{abstract}

Arrays of bioinspired artificial hair-cell airflow velocity sensors can enable flight-by-feel of small, unmanned aircraft. Natural fliers - bats, insects, and birds - have hundred or thousands of velocity sensors distributed across their wings. Aircraft designers do not have this luxury due to size, weight, and power constraints. The challenge is to identify the best locations for a small set of sensors to extract relevant information from the flow field for the prediction of flight control parameters. In this paper, we introduce the data-reducing Sparse Sensor Placement Optimization for Prediction algorithm which locates near-optimal sensor placement on airfoils and wings. For two or more sensors this algorithm finds a set of sensor locations (design point) which predicts angle of attack to within 0.10$\degree$ and ranks within the top 1$\%$ of all possible design points found by brute force search. We demonstrate this algorithm on several variations of airfoil sections of infinite and finite wings in clean and noisy data, evaluate model sensitivities, and show that the algorithm can be used to identify an appropriate number of sensors for a given accuracy requirement. Applications for this algorithm are explored for aircraft design and flight-by-feel control. 

\end{abstract}

\section{Motivation and Contributions}

\begin{figure}
\centering 
\includegraphics[width=3.3in]{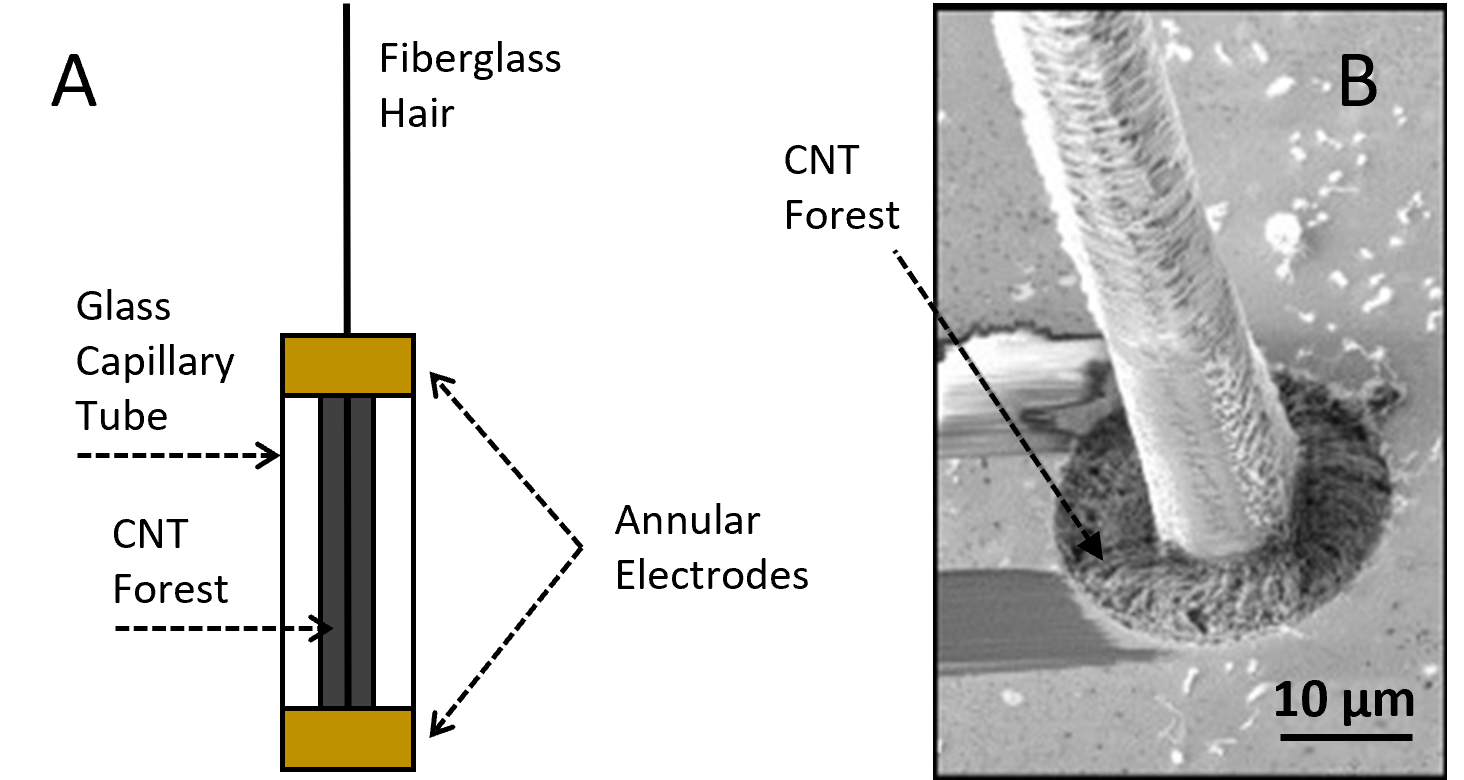}
\caption{A CNT-based AHS. A. Schematic. B. SEM image of hair root and CNT-filled capillary \cite{Phillips2015DetectionSensors}.}
\label{fig:AFRL_AHS}
\end{figure}

Flight-by-feel (FBF) is an emerging paradigm in flight control which may employ integrated, distributed arrays of pressure, strain, and flow sensors to enable rapid and agile flight state estimation and response. A FBF approach may be lighter, faster, more robust, and capture more information than a conventional control system \cite{Salowitz2013Bio-inspiredComposites}. These systems might be comprised entirely or partially of biologically-inspired artificial hair-cell flow sensors (AHS), which detect velocity magnitude from mechanical drag force on hair-like structures. An example of an isotropic (non-directional) AHS is shown in Figure \ref{fig:AFRL_AHS}. This AHS consists of a glass microcapillary tube with an alumina-coated structural glass fiber. A carbon nanotube (CNT) forest is grown inside the capillary. Bending of the exposed hair compresses the CNTs and causes a piezoresistive change between electrodes on the microcapillary tube \cite{Slinker2016CNT-BasedSensing, Maschmann2014Bioinspireddetection}. A comprehensive 2023 survey by the authors evaluates this and other recent AHS designs to synthesize sensor design, function, placement optimization, FBF control, and next-generation aircraft design into a cohesive research paradigm \cite{Hollenbeck2023BioinspiredReviewb}. 

Flying animals have dozens to thousands of hair-cells distributed across their wings. The extent of optimization and the relation to flight control in biological fliers are complex given the vast set of constraints and the processing of multimodal signals as one composite signal \cite{Aiello2021SpatialPhylogeny}. Similar constraints motivate optimized sensor locations in artificial FBF systems. Sparsity is driven by size, weight, and power (SWaP) reduction, potential viscous coupling problems, control actuator design, and other placement considerations \cite{krijnen2019Insect-inspiredsensors, Khot1992EffectStructures}. SWaP restrictions are especially applicable at the micro- or small-UAS scale. Therefore, selecting optimal locations for a given budget of flow sensors is a primary challenge for designing FBF-enabled aircraft. The present research evaluates a new algorithm for near-optimal placement of flow sensors for flight state prediction for implementation on arbitrary airfoils. 

The Sparse Sensor Placement Optimization for Prediction (SSPOP) algorithm presented here is an adaptation of the Sparse Sensor Placement Optimization for Classification (SSPOC) algorithm developed by Brunton et al. \cite{Brunton2016SparseClassification} and later expanded for Reconstruction (SSPOR) by the same group \cite{Mohren2018Neural-inspiredData, Manohar2018IEEE, Callaham2019RobustRepresentation}. The classification approach selects a category for a given data set from among pre-determined classes, based on sparse sampling. The reconstruction approach aims to recreate or model a high-order data set (such as an airflow field) from sparse sampling. The prediction approach differs in that the algorithm's design point is used to calculate or predict a parameter, in this case angle of attack, in a continuous manner rather than select it from a predefined set. These optimization algorithms all use data reduction techniques to reduce the dimensionality of high-order systems. Like artificial hair sensors, data reduction in the context of sensing may be considered bioinspired; animals interact with high-dimensional physical systems via limited sensory information, a form of compressive sensing of big data \cite{Brunton2013RatsDecision-making, bright2013}. In the present research, we use data-driven methods to find near-optimal AHS placement on an airfoil of an infinite wing (Phase I) and a chordwise section of a finite wing (Phase II). Airflow data was computed by linear panel method (Phase I) and CFD (Phase II). A 45$\degree$ sweep delta wing was used for Phase II to evaluate the SSPOP algorithm's effectiveness in complex flow conditions. 

The algorithm specifies a set of sensor locations, called the design point (DP), for a given flow dataset and number of sensors \textit{Q}. This solution is not a true optimum or best possible sensor arrangement. The algorithm selects a DP based on the information features of the flow dataset which is independent of its application (in this case, linear regression of flow measurements at the DP for predicting the angle of attack $\alpha$). To determine its proximity to optimality, the performance of the SSPOP DP in terms of predictive accuracy was evaluated by comparison with the best possible DP found by a brute force search of all possible DPs for small \textit{Q} values. Modifications to the airfoil models were analyzed to determine the sensitivity of input data and model design regarding SSPOP performance. Prediction of other flight metrics based on the SSPOP DP were also investigated. This work represents the first application of a data-driven approach for optimal flow sensor placement on airfoils. The SSPOP algorithm is shown to be highly flexible in scale and scope, with promising FBF implications for sensors of any type and aircraft of any size.   

\section{Methodology - Sparse Sensor Placement Optimization for Prediction}

Each airfoil model includes a set of nodes, or candidate AHS locations, over a 2D surface at which velocity magnitude is known by calculation. The local velocity magnitude at each node represents the theoretical measurement of an AHS placed at that node, assuming perfect fidelity in sensor signals and ignoring physical mounting considerations or boundary layer size. Nodes can be considered binary variables. For any given DP with \textit{Q} sensors and \textit{N} nodes, there will be \textit{Q} local measurements and \(N-Q\) nodes with null measurements; the DP surface velocity distribution is sparse while the full data set is semi-continuous. When a linear regression is performed on the variation of velocity at each DP node over a range of conditions, the prediction model will have \textit{Q} nonzero coefficients and ${N-Q}$ coefficients of zero value. The solution space poses a combinatorial problem: for any given \textit{Q} and \textit{N} there are \(\frac{N!}{Q!(N-Q)!}\) possible DPs.

\subsection{Phase I - Airfoil Models}

\begin{figure}
\centering
\includegraphics[width=6.35in]{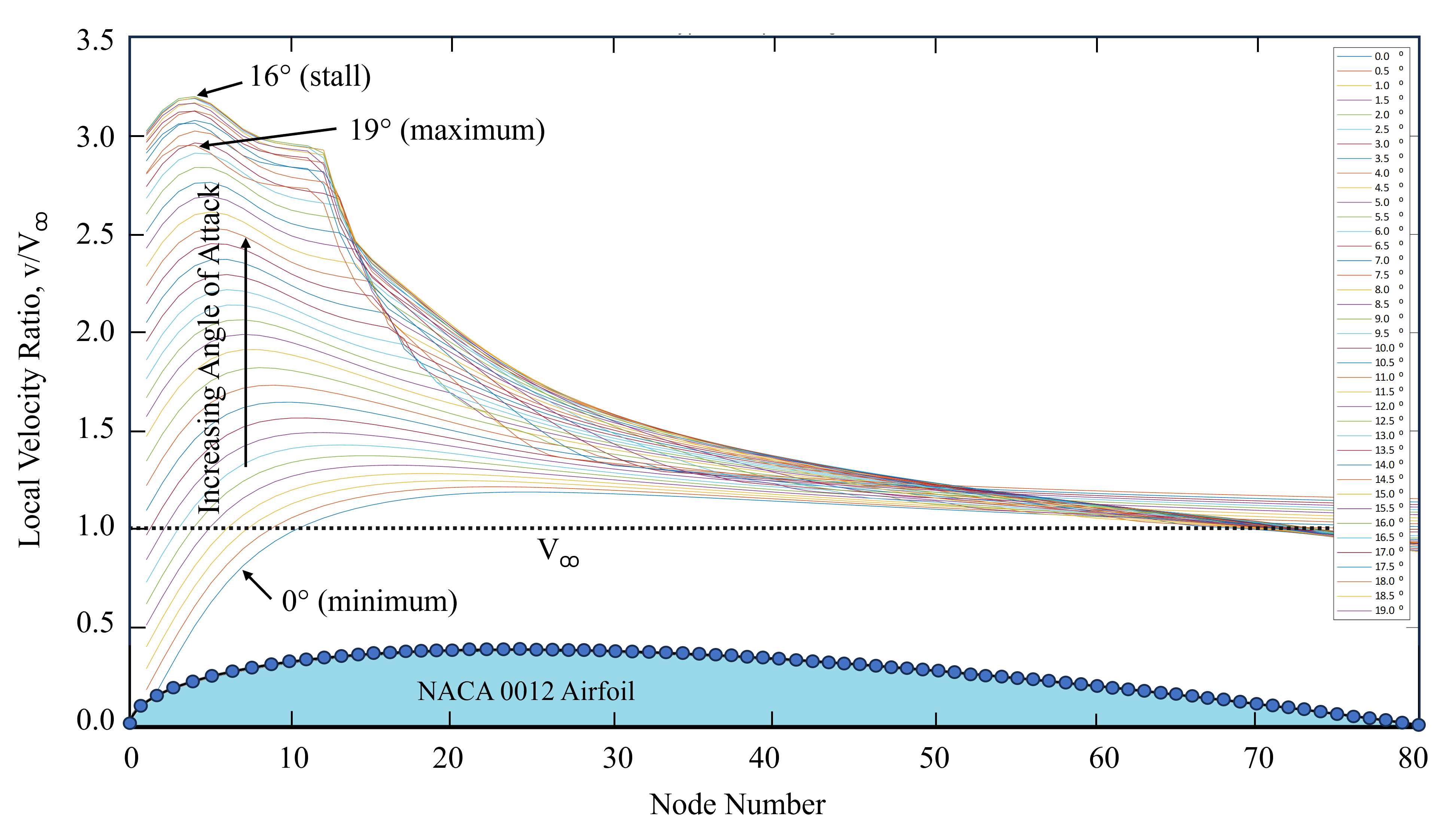}
\caption{Velocity profiles for the Phase I airfoil.}
\label{fig:NACA0012_4}
\end{figure}

The SSPOP algorithm was first applied to a NACA 0012 in a freestream velocity ($V_{\infty}$) of 20 m/s (Re 700,000). The upper surface was divided into 79 panels (Figure \ref{fig:NACA0012_4}) and the local near-surface tangential velocity magnitudes were calculated in XFOIL at $\alpha$ from zero to 19$\degree$ in half-degree increments. XFOIL uses a 2D panel method to calculate the static pressures at panel intersections (nodes) and from this extracts aerodynamic coefficients and velocity information \cite{Drela1989XFOIL:Airfoils}. Flow features such as the transition point are readily apparent at higher $\alpha$. Subsequent iterations of the airfoil model included both the top and bottom surfaces, variations of node number and location constraints, and airfoil shape. Data collection and organization was performed by XFOIL for all Phase I models.

\subsection{Phase II - Delta Wing Slice} \label{CFD}

The model for Phase II was a NACA 4415 delta wing with 45 degree sweep and a root chord of 250 mm (shown in Section \ref{Phase2}, Figure \ref{fig:PhaseIIprofiles3}). The wing geometry was constructed in Ansys DesignModeler as a half-span wing in a tunnel setting. Ansys Workbench and Ansys Fluent were used to acquire velocity data over a wide range of $\alpha$ at 10 m/s $V_{\infty}$. Ansys Fluent was used for grid generation of a polyhedral volume mesh limited to approximately 500 thousand cells for efficiency. Turbulent steady-state estimations of the flow field at 10 m/s were solved for each $\alpha$ using constant density gas at standard conditions with a two-equation k-$\omega$ Shear-Stress Transport model. Each case was solved using a pseudo-transient pressure-based solver where the momentum equation and pressure-based continuity equations are solved simultaneously. This approach was validated by modeling a similar wing matching the geometry and conditions of an experimental wind tunnel test from \cite{Demoret2020TheNumber}, finding errors of approximately 2$\%$ for \(c_l\) and \(c_d\) and 0.20$\%$ for L/D ratio. 

Velocity magnitude values were calculated at 2 mm above the surface around a chordwise slice of the wing at mid-half-span, extracting a 2D airfoil velocity profile from the 3D flow dataset. Velocity magnitudes at each node over this wing slice over a wide range of angles of attack were fed into the SSPOP algorithm to find optimal AHS placements. We then draw conclusions about the performance of the SSPOP algorithm considering the 3D nature of the flow in Phase II compared with the 2D flow in Phase I.

\subsection{SSPOP Algorithm Design}

\begin{figure}
\centering 
\includegraphics[width=\textwidth]{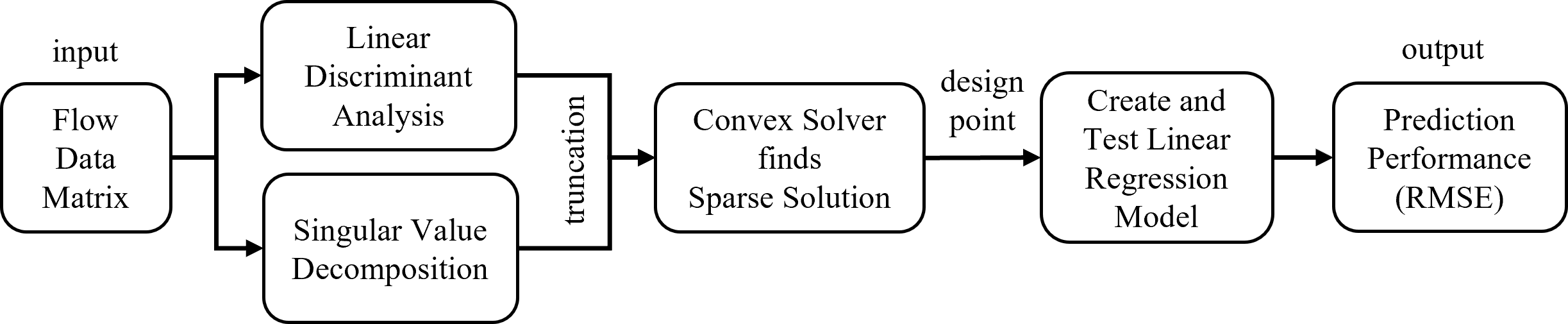}
\caption{The Sparse Sensor Placement Optimization for Prediction (SSPOP) algorithm.}
\label{fig:simpleSSPOPflow}
\end{figure}

Dimensionality reduction by singular value decomposition (SVD) and linear discriminant analysis (LDA) extracts information from large data sets with sparse sampling. The SSPOP algorithm (Figure \ref{fig:simpleSSPOPflow}) uses these methods to identify a sparse set of sensor locations on an airfoil for predicting $\alpha$ from flow data. It begins with SVD of the computationally-obtained near-surface velocity magnitude data to identify prominent flow field features. The data is then truncated using a weights matrix via LDA. The highly truncated data sets contain well over 95$\%$ of the information of the full dataset. A convex solver then creates the final sensor matrix \textit{s}, from which the DP is extracted. Finally, linear regression of AHS data at the DP is performed to obtain a root mean square error (RMSE) of prediction for $\alpha$. This RMSE between the simulation and the sparsely calculated linear regression prediction is the performance metric of the DP. Iteration may occur at any point where improvement is desired. For example, the number of sensors \textit{Q} may be increased until a desired predictive performance is achieved. See Section \ref{repro} for a link to supplementary information including the Matlab code and detailed flow chart for this algorithm.

\subsubsection{Singular Value Decomposition}

The SVD decomposes the training data matrix \textit{X} into three components: a left singular vector matrix \textit{U}, a sparse diagonal matrix of singular values $\Sigma$, and a matrix of right singular vectors \textit{V}. The columns of \textit{U} and \textit{V} are the eigenmodes, features, or components of the data arranged by prominence in terms of being able to describe the variance in \textit{X}. The relative significance of each feature is given by its singular value in $\Sigma$. The singular value spectrum can be used to select a truncation factor \textit{r}. It is desirable to select the lowest \textit{r} which retains a large enough portion of the information in \textit{X}. The importance of the first few columns of \textit{U} and their corresponding singular values and rows of \textit{V} greatly outweigh the remaining several columns. We can drop all but the \textit{r}-most prominent features of the model to create a low-rank reconstruction of \textit{X} with good fidelity. The result of the SVD is an approximation $\tilde{X}$ that is nearly equivalent to \textit{X} in terms of information. The decomposition is essentially a summation of \textit{m} full matrices of size \textit{X}, with the first few terms in the summation making up the majority of the magnitude of the elements of \textit{X}. 

In summary, the SVD extracts prominent features from the high-order full model \(X=U*\Sigma *V^{T}\) and creates a low-order approximation \(\tilde{X}=\Psi *\tilde{\Sigma}*V^{*^{T}}\) which retains most of the original information.

\subsubsection{Linear Discriminant Analysis}

LDA forms a weights matrix based on the variances of the original data matrix \textit{X}.  In general, the objective in classification by LDA is to maximize the ratio of the between-class variance to the within-class variance. The velocity data in the present study has been obtained computationally, so the within-class variance over $\alpha$ in local velocity at each node is zero. Therefore, the LDA in this case simply maximizes the between-class variances. The resulting matrix \textit{W} is normalized and then truncated to the topmost \textit{r} rows. If data is gathered in a non-deterministic manner, for example, with experimental measurements, then repeat measurements at a given $\alpha$ would result in within-class variance. 

\subsubsection{Locating the Design Point}
\label{MethodsDP}

The final step to identify the optimal AHS location is to use a convex solver to find a matrix \textit{s} such that

\begin{center}
$s=argmin_s' \left \{ || s' ||_1+\lambda || s'v ||_1 \right \}$
\end{center}

subject to 

\begin{center}
$|| \Psi_r^T s' - w ||_F \leq \varepsilon $
\end{center}

We seek matrix \textit{s} with a minimum one-norm while keeping the Frobenius norm \(|M|_F=\sqrt{{\sum}_{ij}|m_{ij}|^{2}}\) of the error between the truncated weights and features matrices (\textit{W} and $\Psi$) within some very small tolerance $\varepsilon$. The weights coupling term $\lambda$ determines how strongly the columns of \textit{s} are coupled. When \(\lambda = 0\), the solution for \textit{s} is obtained by an uncoupled, independent approach. As $\lambda$ increases, the columns of \textit{s} become coupled so the same node can now contribute to multiple linear discriminant projection vectors. $\lambda$ and the truncation factor \textit{r} are selected by investigation. To solve for \textit{s}, we used CVX, a MATLAB package for specifying and solving convex programs \cite{Grant2008GraphPrograms, Grant2020}. The \textit{s} matrix is the size of the original data matrix. The DP for any number of sensors \textit{Q} is found by locating the \textit{Q} rows with the \textit{Q}-highest magnitude element values. These rows correspond to the nodes which are “switched on” in the model for optimal information. The data at these nodes will be retained while all remaining nodes will be “switched off” for the next step: linear regression of the DP.

\subsection{Linear Regression} \label{linreg}

Once a DP is selected, a regression is performed on the sparse data matrix $\Psi$ to obtain a linear regression model with coefficients \(\beta_i\) where \(1\le{}i\le{}N\). Only \textit{Q} of these coefficients are nonzero. This equation is then used to calculate the predicted $\alpha$ using the velocity data from the test set, and the results are compared against the actual $\alpha$ to evaluate the performance of the regression model by the RMSE. Unless otherwise specified, all references to "RMSE" in this paper refer to the error of prediction of $\alpha$ in degrees.

\subsection{Validation of SSPOP Performance}

The predictive accuracy of a DP is measured by the RMSE in predicting $\alpha$. In general, the predictive accuracy improves with increasing number of sensors \textit{Q} and varies greatly across the thousands or millions of DPs for a given \textit{Q}. An objective maximum RMSE may be applied based on the FBF controller and desired performance. In the present work, no RMSE performance objective is given other than to find the best possible performance subject to the constraints of the solution method. The SSPOP approach can quickly predict the performance of any number of sensors, so an flight control designer could incorporate these metrics into a tradeoff analysis for the number of sensors. After performing SSPOP for a range of \textit{Q}, the DP with the lowest \textit{Q} meeting a control-system-driven predictive accuracy requirement would be considered the optimum for the given criteria.

The SSPOP DP can be ranked against every possible DP by performing a brute force combinatorial regression analysis. Analysis by this method is limited to low \textit{Q}, especially when \textit{N} is large. As stated earlier, for any given \textit{Q} and \textit{N} there are \(\frac{N!}{Q!(N-Q)!}\) possible DPs. For example, for the simple model with 80 nodes the 6-sensor case would have over 300 million DPs, and the 7-sensor case over 1 billion DPs. The problem quickly becomes intractable for even a moderate number of sensors. As will be shown in the Results section, a brute force analysis of one through four or five sensors serves to establish the general performance of the SSPOP algorithm. While the SSPOP rarely finds the best possible DP, all SSPOP DPs were within or near the top 1$\%$ of DPs for all but the trivial one-sensor case. This analysis indicates that it is reasonable to expect the SSPOP DP for higher \textit{Q} values to be among the best-performing DPs for that number of sensors. Finally, the predictive performance of the SSPOP DP can be used to baseline a search of all possible DPs. Results of this "SSPOP+S" approach are discussed in Section \ref{greedy}.

\subsection{Summary of Methodology} \label{methodsum}

Phase I consists of the development and application of the SSPOP algorithm for a 2D airfoil. Phase II employs the same algorithm on a 2D airfoil section of a 3D wing. The flow velocity magnitude in all cases is derived computationally, using XFOIL for 2D velocity data and CFD for 3D velocity data. 

The SSPOP algorithm will always pick the same DP for a given data set and sensor number (and weighting and truncation factors). That is, the algorithm selects the most information-rich sensor locations independent of the application or prediction metric. The DP is only optimal in the sense that it captures very near the greatest possible fraction of the information from the complete dataset. The present work focuses on $\alpha$ prediction. The same analysis can be performed with other aerodynamic parameters such as the coefficients of lift, drag, and moment, which each vary independently with $\alpha$ and other flight conditions. Brief results of such an analysis are presented in Section \ref{multi}. 

\section{Results}

Phase I included two airfoil models: a NACA 0012 top surface-only model, and a NACA 4415 full-surface model. Each included several variations in node placement and number. Several variations in the model set-up and pre-processing of node velocity data were analyzed during Phase I, as part of a full analysis of the performance of the SSPOP algorithm and to develop best practices for its application to FBF. The results of SSPOP performance for Phase II are then given briefly. Section (\ref{diss}) summarizes all results with particular attention to real-world implications. 

\subsection{Phase I.a - NACA 0012 Airfoil}
\label{Phase1aresults}

Table \ref{tab:NACA0012SSPOP} shows the SSPOP DP performance by RMSE in $\alpha$ prediction and ranking among all possible DPs for up to five sensors. The best possible (BP) RMSE found by brute force search is shown for comparison. The ranking of the SSPOP DP against all possible DPs establish confidence in the performance of the SSPOP algorithm for higher values of \textit{Q} and more complex models where the brute force search is not feasible. Section \ref{multi} discusses the trends in predictive accuracy as \textit{Q} increases toward \textit{N}.

\begin{table}
\centering
\small
\caption{Phase I.a results establish confidence in SSPOP performance for predicting $\alpha$.}
\label{tab:NACA0012SSPOP}
\begin{tabular}{
>{\columncolor[HTML]{FFFFFF}}c 
>{\columncolor[HTML]{FFFFFF}}c 
>{\columncolor[HTML]{FFFFFF}}c 
>{\columncolor[HTML]{FFFFFF}}c }
Q                                           & type                       & RMSE, deg               & \multicolumn{1}{l}{\cellcolor[HTML]{FFFFFF}Rank, $\%$} \\ \hline
\cellcolor[HTML]{FFFFFF}                    & SSPOP                      & 2.097                         & 6.25                                                   \\
\multirow{-2}{*}{\cellcolor[HTML]{FFFFFF}1} & \cellcolor[HTML]{EFEFEF}BP & \cellcolor[HTML]{EFEFEF}1.260 &                                                        \\ \hline
\cellcolor[HTML]{FFFFFF}                    & SSPOP                      & 0.427                         & 0.35                                                   \\
\multirow{-2}{*}{\cellcolor[HTML]{FFFFFF}2} & \cellcolor[HTML]{EFEFEF}BP & \cellcolor[HTML]{EFEFEF}0.397 &                                                        \\ \hline
\cellcolor[HTML]{FFFFFF}                    & SSPOP                      & 0.150                         & 1.11                                                   \\
\multirow{-2}{*}{\cellcolor[HTML]{FFFFFF}3} & \cellcolor[HTML]{EFEFEF}BP & \cellcolor[HTML]{EFEFEF}0.106 & \multicolumn{1}{l}{\cellcolor[HTML]{FFFFFF}}           \\ \hline
\cellcolor[HTML]{FFFFFF}                    & SSPOP                      & 0.148                         & 0.48                                                   \\
\multirow{-2}{*}{\cellcolor[HTML]{FFFFFF}4} & \cellcolor[HTML]{EFEFEF}BP & \cellcolor[HTML]{EFEFEF}0.082 & \multicolumn{1}{l}{\cellcolor[HTML]{FFFFFF}}           \\ \hline
\cellcolor[HTML]{FFFFFF}                    & SSPOP                      & 0.127                         & 0.63                                                   \\
\multirow{-2}{*}{\cellcolor[HTML]{FFFFFF}5} & \cellcolor[HTML]{EFEFEF}BP & \cellcolor[HTML]{EFEFEF}0.063 & \multicolumn{1}{l}{\cellcolor[HTML]{FFFFFF}}           \\ \hline
\end{tabular}
\end{table}

\subsubsection{SSPOP plus Search} \label{greedy}

As previously mentioned, the SSPOP DP performance can serve as a starting point for a search algorithm. For example, the best possible DP for 5 sensors had an RMSE of 0.0627$\degree$. SSPOP finds a DP with a 0.1267$\degree$ RMSE, which while only half as good is still within the top 0.63$\%$ of possible DPs. Using the SSPOP DP performance as a target to beat, a search finds a better DP very quickly. Using this method, a DP with an RMSE of 0.1065 $\degree$ was found after only a few minutes. This new DP ranked in the top 0.0072$\%$ of all possible DPs. The SSPOP+S process works the same for cases where the best possible DP is not known, since we can be confident that SSPOP finds a DP in or near the top 1$\%$. 

\subsubsection{Adding Noise}
\begin{figure}[hb]
\centering
\includegraphics[width=\textwidth]{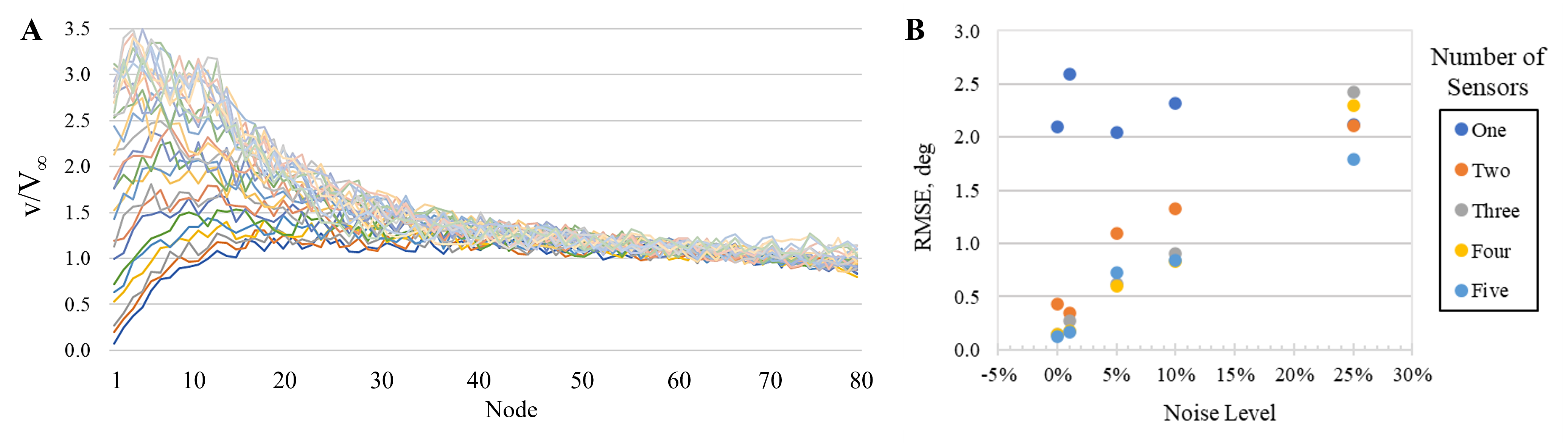}
\caption{A. Velocity profiles with up to 10$\%$ artificial noise. \hspace{.1mm} B. SSPOP DP RMSE for clean data and in artificial noise up to 1, 5, 10, and 25 percent. Note: The one and two sensor RMSE at 25$\%$ noise are equivalent.} 
\label{fig:Noise4}
\end{figure}

The initial analysis used clean computational data, essentially assuming perfect fidelity in AHS signals. True AHS signals are noisy, so artificial noise was added to the XFOIL results to examine its effect of SSPOP performance.  The original data was multiplied by a percentage of a random value between -1 and 1, yielding artificially noisy data which varies randomly up to the specified percent noise in either direction (Figure \ref{fig:Noise4}.A). Comparing with the original data in Figure \ref{fig:NACA0012_4}, flow features such as the transition points which were clearly evident are now obscured in the noisy data. The SSPOP DP was found for 1, 5, 10, and 25 percent noise and compared with the no-noise performance (Figure \ref{fig:Noise4}.B). Predictably, the chart confirms that the SSPOP DP RMSE increases linearly with increasing noise for all but the trivial one-sensor case. Interestingly, the one-sensor performance appears to be independent of noise level. It it possible that this artificial noise analysis overestimates the negative impact of noise. Real-world AHS signals would have non-random noise; for example, greater in turbulent flow and lesser in laminar flow (see Section \ref{turb} below). It is likely that non-random noise levels will contribute useful information to the data set, so the impact of noisy signals on real-world SSPOP applications may be less than modeled here.  

\subsubsection{Sensitivity Analysis: Node Number and Regression Ratio}

\begin{figure}
\centering
\includegraphics[width=5.5in]{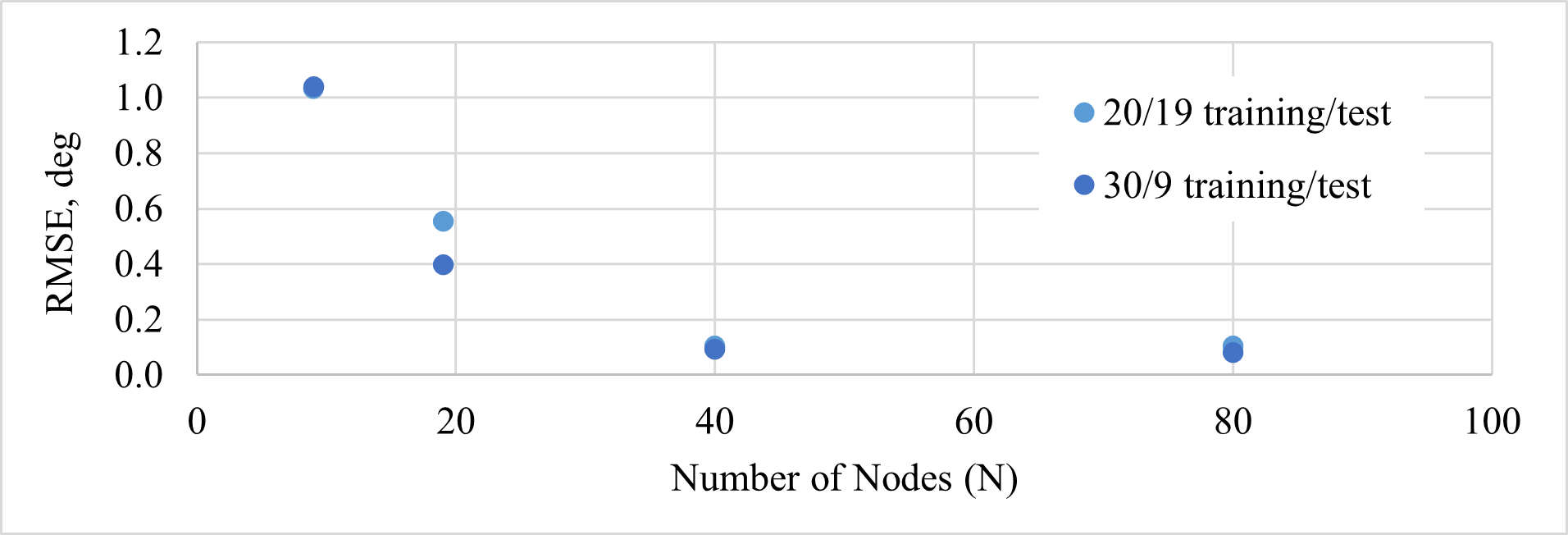}
\caption{Sensitivity analysis of SSPOP performance for \textit{Q} = four sensors for \textit{N} = 9, 19, 40, or 80 and two ratios of training to test runs.} 
\label{fig:SensChart5}
\end{figure}

A sensitivity analysis was performed on the NACA 0012 model to evaluate the impact of number of nodes (\textit{N}) and the ratio of training to test runs for regression building. SSPOP DPs were found for \textit{N} = 9, 19, 39, and 80 nodes with the original 39 runs split into 30/9 and 20/19 training to test ratios. The resulting performances are shown in Figure \ref{fig:SensChart5}. For one through four sensors (only the four-sensor case is shown), the RMSE significantly improved from 9 to 40 nodes but only moderately improved from 40 to 80 nodes, indicating that high density of nodes is not required for good performance. Additionally, using a larger ratio of training runs to test runs improved the RMSE in nearly every case but not significantly at higher node counts. Two conclusions may be derived from this sensitivity analysis: (1) performance gain from denser node spacing is nonlinear; there is a point of diminishing returns where the increased complexity of greater node numbers does not correlate with meaningful gains in performance, and (2) it is best to build the regression model with a larger set of runs. 

\subsubsection{Predicting Additional Aerodynamic Parameters} \label{multi}

\begin{figure}
\centering
\includegraphics[width=\textwidth]{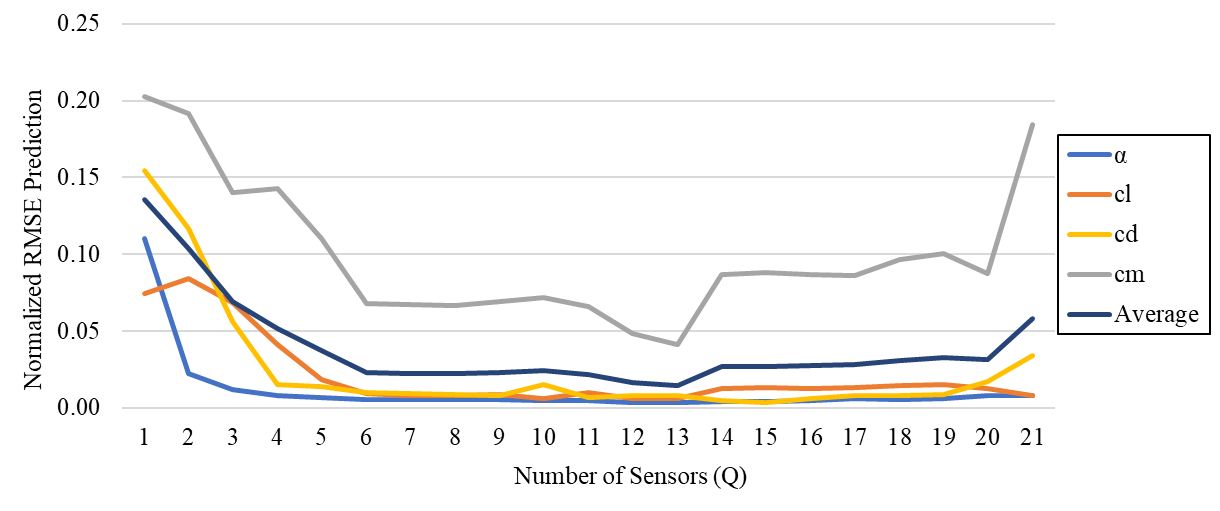}
\caption{Sensitivities for four aerodynamic parameters: $\alpha$, \(c_l\), \(c_d\), and \(c_m\) by RMSE at the SSPOP DP.}
\label{fig:sens}
\end{figure}

As mentioned in Section \ref{methodsum}, the SSPOP algorithm will pick the same DP for a given model regardless of the metric(s) to be predicted. Having shown that the SSPOP DP performs well in predicting $\alpha$, we briefly explored prediction of additional parameters. An investigation of a multi-response problem was performed using the same 80-node NACA 0012 2D model and XFOIL flow data. Values for the aerodynamic coefficients of lift \(c_l\), drag \(c_d\), and longitudinal moment \(c_m\) were obtained for each of the 30 test angles and 9 validation angles. The SSPOP DPs found earlier for each \textit{Q} are still valid for analyzing the performance in predicting these new metrics. A regression model was created for each of the three new metrics and the RMSE for each was found in the same way as for $\alpha$. All metrics were then normalized. Figure \ref{fig:sens} shows that the RMSE of predictions for the aerodynamic coefficients follow the general trend as those for $\alpha$.

Though the SSPOP DP is independent, the best possible DP for each \textit{Q} does vary according to the metric being predicted. The multivariate prediction performance of a DP may be evaluated based on the average normalized RMSE across all four metrics. For this model, the best DP by normalized RMSE average was found by brute search for the four-sensor case. This DP was located at nodes [3 7 27 65] and had an RMSE for each metric within the top 0.30$\%$ of all possible DPs. This demonstrates that in general a DP can be found which accurately predicts multiple metrics.

\subsubsection{Analysis and Summary of NACA 0012}

In Phase I.a we applied the SSPOP algorithm to the upper surface of a symmetric NACA 0012 airfoil to baseline the algorithm's performance and explore model variations. The SSPOP found a near-optimal location for AHSs to predict $\alpha$, \(c_l\), \(c_d\), and \(c_m\) by a sparse linear regression model. A brute force search of all possible DPs for small \textit{Q} values enabled evaluation of SSPOP DP performance. Additionally, an analysis of different model configurations – the same airfoil with 9, 19, 40, or 80 nodes – confirmed the hypothesis that a greater set of candidate sensor positions enabled more accurate predictions for a given \textit{Q}. Finally, evaluation with artificially noisy data demonstrated that SSPOP can extract useful features despite signal noise.

The SSPOP algorithm performed very well for predicting $\alpha$, always finding a DP within the top 1$\%$ of all possible DPs by RMSE. Adding a search procedure to beat the SSPOP performance yielded a DP within the top 0.01$\%$ in a few minutes, versus several to dozens of hours for a brute force search required to find the best possible DP. The SSPOP DP linear regression solution did not perform as well for the other three metrics, especially \(c_l\) and \(c_m\). One possible avenue would be to perform the same procedure using the average normalized RMSE across all four metrics. To that end, the best DP by normalized RMSE average was found by brute search for the four-sensor case. This DP had an RMSE for each metric within the top 0.30$\%$ of all possible DPs, demonstrating that a DP exists which accurately predicts multiple metrics. However, this DP was different from the SSPOP DP, and it is unclear whether such an optimization can be achieved by data reduction methods alone. These findings indicate that the predictive power of the SSPOP DP is dependent on the type and quality of the flow information input to the algorithm.

\subsection{Phase I.b - NACA 4415 Full-Wrap Model} \label{NACA4415}

\begin{figure}[hb]
\centering 
\includegraphics[width=5.75in]{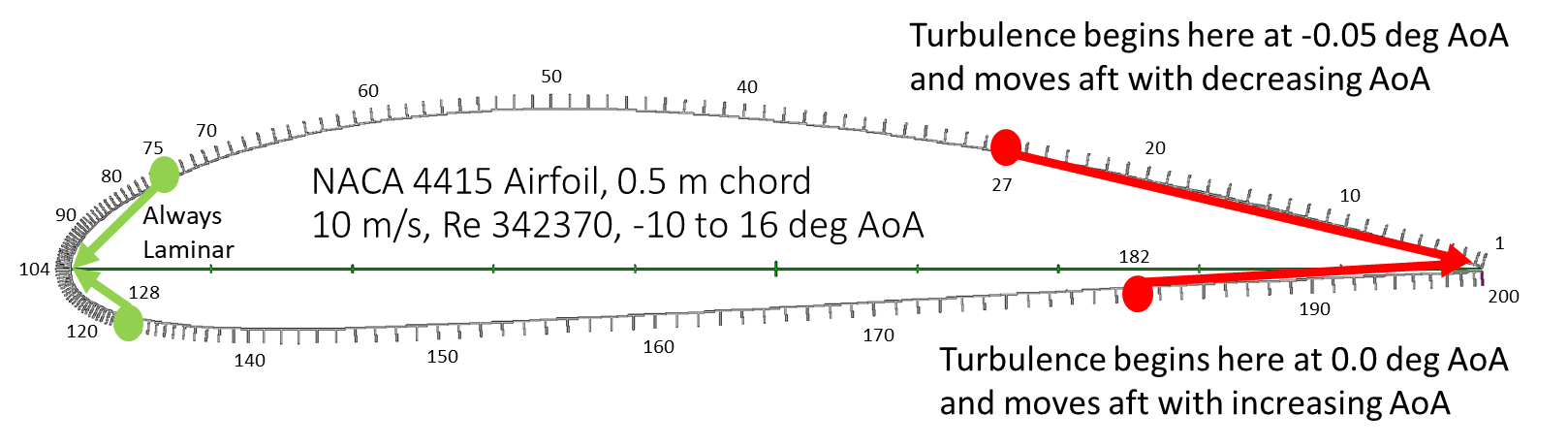}
\caption{Phase I.b model.}
\label{fig:NACA4415}
\end{figure}

Phase I.a served to validate the SSPOP algorithm on a simple model. The next step was to evaluate the SSPOP algorithm on an asymmetric airfoil with nodes wrapped around the entire surface. The surface airflow velocity over a NACA 4415 airfoil was calculated in XFOIL. The airfoil had a chord of 0.5 m and the conditions were sea level standard at 10 m/s freestream. In the initial model for Phase I.b, local velocity magnitude was found at each of 200 nodes over the surface of the airfoil. With this larger model there are over 64 million possible DPs for four sensors. The nodes are numbered starting at 1 at the upper trailing edge (TE) and proceeding counterclockwise around the airfoil to node 200 at the lower TE. The nodes are spaced closer together near the leading edge (LE) for better fidelity in that region of higher pressure and velocity gradients. Data was gathered for $\alpha$ from negative 10$\degree$ to positive 16$\degree$ then divided into 53 cases for model building and 15 cases for validation. Portions of the LE experienced entirely laminar flow throughout the wide range of $\alpha$, as indicated in Figure \ref{fig:NACA4415}. Likewise, portions of the TE experienced entirely turbulent flow. Some variations of the model accounted for the variable laminar-to-turbulent transition point, as discussed below. 

\subsubsection{Turbulent Flow}
\label{turb}

Some AHSs respond cleanly in slower and laminar flow but saturate or get very noisy in turbulent flow \cite{Slinker2016CNT-BasedSensing, Magar2019AerodynamicFlow}. Accounting for these phenomena may be crucial for modeling a realistic response. Referring again to Figure \ref{fig:NACA4415}, only a small portion of the LE is fully laminar over this large $\alpha$ range (-10$\degree$ to 16$\degree$), but sensors on the top and bottom will experience mostly laminar flow in negative and positive $\alpha$, respectively. By instrumenting a wing with sensors on the top and bottom in equal portions, and avoiding areas near the TE, approximately half of the sensors will experience laminar flow at any given flight state. This indicates that top/bottom sensor redundancy may be desirable in FBF systems.

\begin{figure}
\centering 
\includegraphics[width=6.3in]{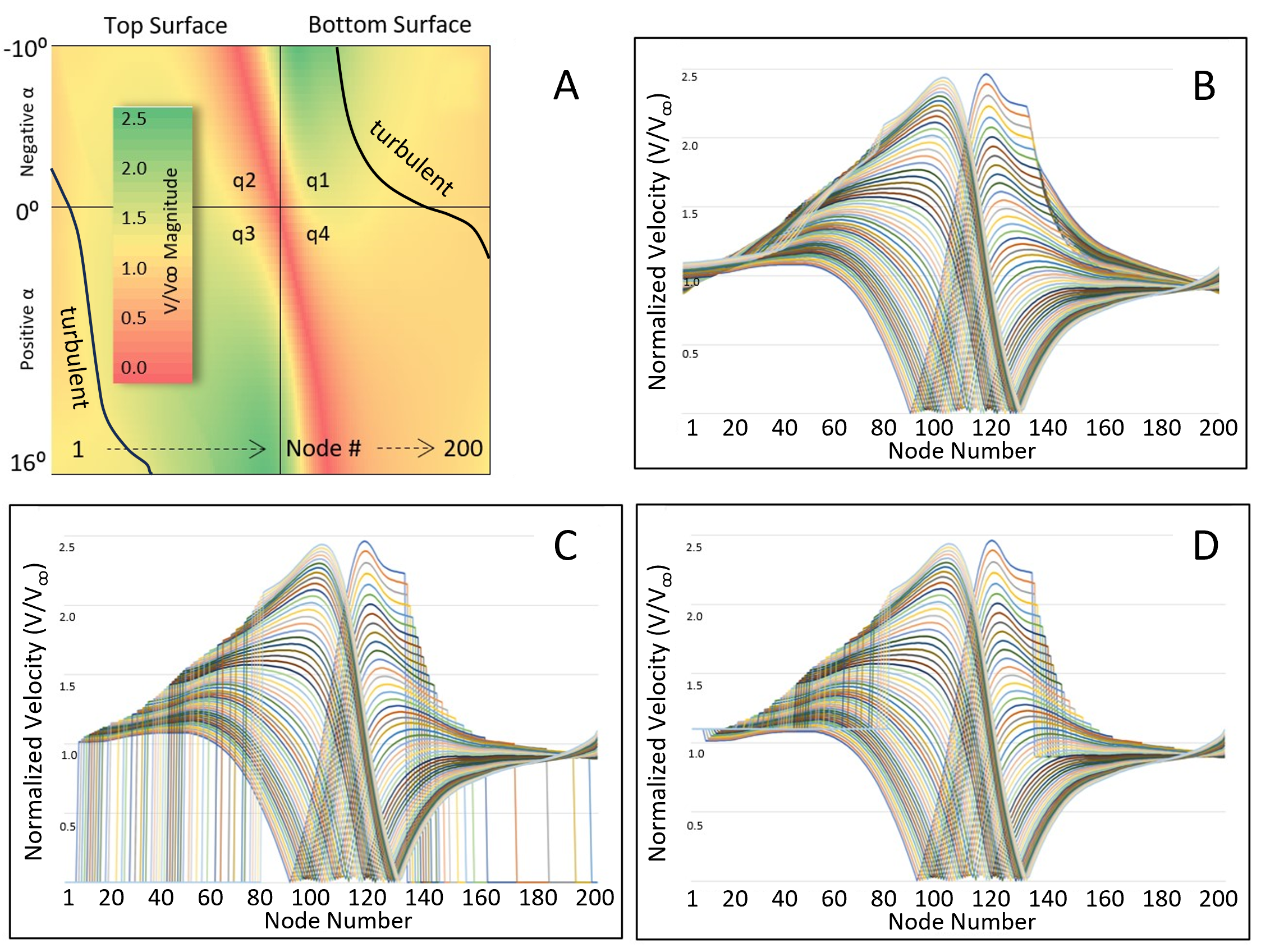}
\caption{Accounting for Turbulence. \hspace{.3mm} A. Original data matrix shaded by velocity ratio magnitude. \hspace{.3mm} B. Velocity profiles, each line is one $\alpha$. \hspace{.3mm} C. Turbulent data = zero. \hspace{.3mm} D. Turbulent data = local ave.: 1.1 (top) and 0.9 (bottom).}
\label{fig:TurbPlots}
\end{figure}

Figure \ref{fig:TurbPlots} depicts the original data for Phase I.b and two variations with turbulent adjustments. Figure \ref{fig:TurbPlots}.A shows the full set of flow data at each node for each $\alpha$ (size 53 x 200 matrix), shaded by flow magnitude. The same data is graphed in Figure \ref{fig:TurbPlots}.B with one line for each $\alpha$. Some flow features are apparent: The stagnation point migrates around the LE with changing $\alpha$. The low pressure region moves from top to bottom as $\alpha$ changes from positive to negative. The turbulent transition front is outlined on the matrix in Figure \ref{fig:TurbPlots}.A and is apparent in the sharp changes in velocity in Figure \ref{fig:TurbPlots}.B. Figure \ref{fig:TurbPlots}.C shows the velocity profiles with all turbulent data overwritten to zero. This is representative of "turning off" the sensors when turbulent data is detected. Figure \ref{fig:TurbPlots}.D shows the data with turbulent values tuned to the approximate average of the calculated normalized velocities in that region: 1.1 for the top surface and 0.9 for the bottom.

\begin{figure}
\centering 
\includegraphics[width=\textwidth]{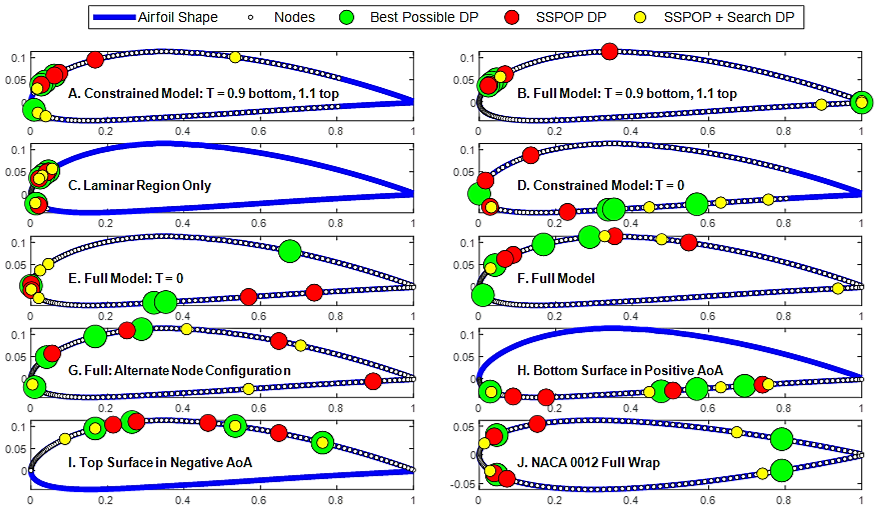}
\caption{Phase I.b: 4-sensor DPs for all model variations. Subplot labels correspond to Table \ref{tab:Results4415_2}. Colored dots show location of sensors for each DP. Nodes or candidate sensor locations are represented by white dots.}
\label{fig:PhaseIplots3}
\end{figure}

The data matrices in Figure \ref{fig:TurbPlots} are divided into quadrants: q1 is the bottom surface in negative $\alpha$, q2 is the top surface in negative $\alpha$, q3 is the bottom surface in positive $\alpha$, and q4 is the top surface in positive $\alpha$. Quadrants q1 and q3 are largely turbulent, while q2 and q4 are largely laminar. Based on insights from the Phase I.a NACA 0012 analyses and the features of this new dataset, several variations of the model were analyzed in terms of best possible DP performance in $\alpha$ prediction, absolute SSPOP DP and SSPOP+S DP performance, and relative (SSPOP vs best possible) performance. These variations and their results are described in Table \ref{tab:Results4415_2} and Figure \ref{fig:PhaseIplots3}.

\subsubsection{Model Variations}

The purposes of model variation include: (1) Characterizing the performance of the SSPOP algorithm subject to various physical and data constraints. (2) Analyzing the impact of model construction on SSPOP performance. (3) Analyzing performance of linear regression in predicting $\alpha$ from sparse sensor measurements subject to various constraints. (4) Characterizing the relationships between absolute and relative performance of SSPOP DPs by RMSE. In all cases, brute force searches were performed for up to four sensors to validate the SSPOP algorithm. \\

\begin{table}
\caption{Phase I.b results for 9 variants of the NACA 4415 airfoil model (A-I) and a full-surface variant of the NACA 0012 airfoil from Phase I.a (J). All DP locations are given in full model node numbering equivalent.}
\label{tab:Results4415_2}
\resizebox{\textwidth}{!}{%
\begin{tabular}{@{}
>{\columncolor[HTML]{FFFFFF}}c 
>{\columncolor[HTML]{FFFFFF}}c 
>{\columncolor[HTML]{FFFFFF}}r 
>{\columncolor[HTML]{FFFFFF}}c 
>{\columncolor[HTML]{FFFFFF}}c 
>{\columncolor[HTML]{FFFFFF}}c 
>{\columncolor[HTML]{FFFFFF}}c 
>{\columncolor[HTML]{FFFFFF}}c 
>{\columncolor[HTML]{FFFFFF}}c 
>{\columncolor[HTML]{FFFFFF}}c 
>{\columncolor[HTML]{FFFFFF}}c 
>{\columncolor[HTML]{FFFFFF}}r 
>{\columncolor[HTML]{FFFFFF}}c 
>{\columncolor[HTML]{FFFFFF}}c 
>{\columncolor[HTML]{FFFFFF}}c 
>{\columncolor[HTML]{FFFFFF}}c 
>{\columncolor[HTML]{FFFFFF}}c 
>{\columncolor[HTML]{FFFFFF}}c @{}}
\toprule
\multicolumn{2}{c}{\cellcolor[HTML]{FFFFFF}Model}                                                                                                                       & \multicolumn{1}{c}{\cellcolor[HTML]{FFFFFF}Method} & \multicolumn{4}{c}{\cellcolor[HTML]{FFFFFF}4-Sensor DP} & \begin{tabular}[c]{@{}c@{}}RMSE\\ (deg $\alpha$)\end{tabular} & \begin{tabular}[c]{@{}c@{}}Rank\\ ($\%$)\end{tabular} & \multicolumn{2}{c}{\cellcolor[HTML]{FFFFFF}Model}                                                                                                                            & \multicolumn{1}{c}{\cellcolor[HTML]{FFFFFF}Method} & \multicolumn{4}{c}{\cellcolor[HTML]{FFFFFF}4-Sensor DP} & \begin{tabular}[c]{@{}c@{}}RMSE\\ (deg $\alpha$)\end{tabular} & \begin{tabular}[c]{@{}c@{}}Rank\\ ($\%$)\end{tabular} \\ \midrule
\cellcolor[HTML]{FFFFFF}                               & \cellcolor[HTML]{FFFFFF}                                                                                       & BP                                                 & 76           & 81           & 84          & 115         & 0.093                                                         & \multicolumn{1}{l|}{\cellcolor[HTML]{FFFFFF}}         & \cellcolor[HTML]{FFFFFF}                               & \cellcolor[HTML]{FFFFFF}                                                                                            & BP                                                 & 54           & 64           & 80          & 116         & 0.038                                                         & \multicolumn{1}{l}{\cellcolor[HTML]{FFFFFF}}          \\
\cellcolor[HTML]{FFFFFF}                               & \cellcolor[HTML]{FFFFFF}                                                                                       & SSPOP                                              & 64           & 74           & 76          & 84          & 0.265                                                         & \multicolumn{1}{c|}{\cellcolor[HTML]{FFFFFF}0.062}    & \cellcolor[HTML]{FFFFFF}                               & \cellcolor[HTML]{FFFFFF}                                                                                            & SSPOP                                              & 35           & 49           & 72          & 75          & 0.140                                                         & 1.488                                                 \\
\multirow{-3}{*}{\cellcolor[HTML]{FFFFFF}\textbf{(A)}} & \multirow{-3}{*}{\cellcolor[HTML]{FFFFFF}\textbf{Constrained     Turb. = AVE}}                                 & SSPOP+S                                            & 36           & 88           & 120         & 127         & 0.243                                                         & \multicolumn{1}{c|}{\cellcolor[HTML]{FFFFFF}0.019}    & \multirow{-3}{*}{\cellcolor[HTML]{FFFFFF}\textbf{(F)}} & \multirow{-3}{*}{\cellcolor[HTML]{FFFFFF}\textbf{\begin{tabular}[c]{@{}c@{}}Full\\      Model\end{tabular}}}        & SSPOP+S                                            & 40           & 51           & 83          & 194         & 0.073                                                         & 0.010                                                 \\ \midrule
\cellcolor[HTML]{FFFFFF}                               & \cellcolor[HTML]{FFFFFF}                                                                                       & BP                                                 & 79           & 82           & 85          & 200         & 0.051                                                         & \multicolumn{1}{l|}{\cellcolor[HTML]{FFFFFF}}         & \cellcolor[HTML]{FFFFFF}                               & \cellcolor[HTML]{FFFFFF}                                                                                            & BP                                                 & 54           & 64           & 80          & 116         & 0.038                                                         & \multicolumn{1}{l}{\cellcolor[HTML]{FFFFFF}}          \\
\cellcolor[HTML]{FFFFFF}                               & \cellcolor[HTML]{FFFFFF}                                                                                       & SSPOP                                              & 50           & 75           & 84          & 85          & 0.197                                                         & \multicolumn{1}{c|}{\cellcolor[HTML]{FFFFFF}0.200}    & \cellcolor[HTML]{FFFFFF}                               & \cellcolor[HTML]{FFFFFF}                                                                                            & SSPOP                                              & 28           & 57           & 77          & 191         & 0.131                                                         & 0.850                                                 \\
\multirow{-3}{*}{\cellcolor[HTML]{FFFFFF}\textbf{(B)}} & \multirow{-3}{*}{\cellcolor[HTML]{FFFFFF}\textbf{Full     Turb. = AVE}}                                        & SSPOP+S                                            & 1            & 77           & 191         & 200         & 0.115                                                         & \multicolumn{1}{c|}{\cellcolor[HTML]{FFFFFF}0.043}    & \multirow{-3}{*}{\cellcolor[HTML]{FFFFFF}\textbf{(G)}} & \multirow{-3}{*}{\cellcolor[HTML]{FFFFFF}\textbf{\begin{tabular}[c]{@{}c@{}}Alt. Node\\      Config.\end{tabular}}} & SSPOP+S                                            & 24           & 45           & 112         & 170         & 0.089                                                         & 0.049                                                 \\ \midrule
\cellcolor[HTML]{FFFFFF}                               & \cellcolor[HTML]{FFFFFF}                                                                                       & BP                                                 & 79           & 82           & 85          & 118         & 0.066                                                         & \multicolumn{1}{l|}{\cellcolor[HTML]{FFFFFF}}         & \cellcolor[HTML]{FFFFFF}                               & \cellcolor[HTML]{FFFFFF}                                                                                            & BP                                                 & 124          & 164          & 170         & 178         & 0.009                                                         & \multicolumn{1}{l}{\cellcolor[HTML]{FFFFFF}}          \\
\cellcolor[HTML]{FFFFFF}                               & \cellcolor[HTML]{FFFFFF}                                                                                       & SSPOP                                              & 79           & 80           & 86          & 121         & 0.156                                                         & \multicolumn{1}{c|}{\cellcolor[HTML]{FFFFFF}0.309}    & \cellcolor[HTML]{FFFFFF}                               & \cellcolor[HTML]{FFFFFF}                                                                                            & SSPOP                                              & 136          & 144          & 166         & 181         & 0.026                                                         & 0.828                                                 \\
\multirow{-3}{*}{\cellcolor[HTML]{FFFFFF}\textbf{(C)}} & \multirow{-3}{*}{\cellcolor[HTML]{FFFFFF}\textbf{\begin{tabular}[c]{@{}c@{}}Laminar\\      Only\end{tabular}}} & SSPOP+S                                            & 77           & 83           & 86          & 117         & 0.067                                                         & \multicolumn{1}{c|}{\cellcolor[HTML]{FFFFFF}0.001}    & \multirow{-3}{*}{\cellcolor[HTML]{FFFFFF}\textbf{(H)}} & \multirow{-3}{*}{\cellcolor[HTML]{FFFFFF}\textbf{\begin{tabular}[c]{@{}c@{}}Bottom\\      Only\end{tabular}}}       & SSPOP+S                                            & 125          & 162          & 174         & 182         & 0.011                                                         & 0.004                                                 \\ \midrule
\cellcolor[HTML]{FFFFFF}                               & \cellcolor[HTML]{FFFFFF}                                                                                       & BP                                                 & 104          & 155          & 156         & 170         & 0.056                                                         & \multicolumn{1}{l|}{\cellcolor[HTML]{FFFFFF}}         & \cellcolor[HTML]{FFFFFF}                               & \cellcolor[HTML]{FFFFFF}                                                                                            & BP                                                 & 20           & 36           & 56          & 64          & 0.001                                                         & \multicolumn{1}{l}{\cellcolor[HTML]{FFFFFF}}          \\
\cellcolor[HTML]{FFFFFF}                               & \cellcolor[HTML]{FFFFFF}                                                                                       & SSPOP                                              & 67           & 88           & 124         & 148         & 0.544                                                         & \multicolumn{1}{c|}{\cellcolor[HTML]{FFFFFF}0.430}    & \cellcolor[HTML]{FFFFFF}                               & \cellcolor[HTML]{FFFFFF}                                                                                            & SSPOP                                              & 28           & 41           & 55          & 60          & 0.009                                                         & 27.93                                                 \\
\multirow{-3}{*}{\cellcolor[HTML]{FFFFFF}\textbf{(D)}} & \multirow{-3}{*}{\cellcolor[HTML]{FFFFFF}\textbf{Constrained     Turb. = 0}}                                   & SSPOP+S                                            & 31           & 104          & 163         & 164         & 0.232                                                         & \multicolumn{1}{c|}{\cellcolor[HTML]{FFFFFF}0.012}    & \multirow{-3}{*}{\cellcolor[HTML]{FFFFFF}\textbf{(I)}} & \multirow{-3}{*}{\cellcolor[HTML]{FFFFFF}\textbf{\begin{tabular}[c]{@{}c@{}}Top\\      Only\end{tabular}}}          & SSPOP+S                                            & 20           & 36           & 64          & 72          & 0.001                                                         & 0.001                                                 \\ \midrule
\cellcolor[HTML]{FFFFFF}                               & \cellcolor[HTML]{FFFFFF}                                                                                       & BP                                                 & 26           & 103          & 154         & 156         & 0.030                                                         & \multicolumn{1}{l|}{\cellcolor[HTML]{FFFFFF}}         & \cellcolor[HTML]{FFFFFF}                               & \cellcolor[HTML]{FFFFFF}                                                                                            & BP                                                 & 15           & 63           & 98          & 146         & 0.033                                                         & \multicolumn{1}{l}{\cellcolor[HTML]{FFFFFF}}          \\
\cellcolor[HTML]{FFFFFF}                               & \cellcolor[HTML]{FFFFFF}                                                                                       & SSPOP                                              & 101          & 109          & 170         & 181         & 0.555                                                         & \multicolumn{1}{c|}{\cellcolor[HTML]{FFFFFF}1.289}    & \cellcolor[HTML]{FFFFFF}                               & \cellcolor[HTML]{FFFFFF}                                                                                            & SSPOP                                              & 54           & 64           & 97          & 101         & 0.119                                                         & 0.819                                                 \\
\multirow{-3}{*}{\cellcolor[HTML]{FFFFFF}\textbf{(E)}} & \multirow{-3}{*}{\cellcolor[HTML]{FFFFFF}\textbf{Full     Turb. = 0}}                                          & SSPOP+S                                            & 79           & 85           & 108         & 121         & 0.439                                                         & \multicolumn{1}{c|}{\cellcolor[HTML]{FFFFFF}0.546}    & \multirow{-3}{*}{\cellcolor[HTML]{FFFFFF}\textbf{(J)}} & \multirow{-3}{*}{\cellcolor[HTML]{FFFFFF}\textbf{\begin{tabular}[c]{@{}c@{}}NACA 0012\\      Wrapped\end{tabular}}} & SSPOP+S                                            & 22           & 70           & 95          & 143         & 0.080                                                         & 0.031                                                 \\ \bottomrule
\end{tabular}%
}
\end{table}

\begin{enumerate}[label=(\Alph*)] %change to \alph
    \item{\textbf{Constrained Model with Turb. = AVE.} The constrained model retains 121 of the original 200 nodes. Many LE nodes were removed to facilitate realistic spacing, and the aft 20$\%$ of the nodes were removed to account for a notional control surface. Turbulent readings were overwritten to the approximate average value of turbulent data points on that side of the airfoil: 1.1 for the top surface and 0.9 for the bottom surface.}
 
    \item{\textbf{Full Model with Turb. = Average.} Same as (A) but with the full 200-node model (\ref{fig:TurbPlots}.D}
    
    \item{\textbf{Laminar Region Only.} The model retains only the 54 nodes from the full model near the LE that are laminar throughout the entire range of $\alpha$ from -10$\degree$ to 16$\degree$.}

    \item{\textbf{Constrained Model with Turb. = 0.} Same constraints as (A) but with all turbulent data points set to zero.}

    \item{\textbf{Full Model with Turb. = 0.} 200 node original model with all turbulent values set to zero (Figure \ref{fig:TurbPlots}.C).}

    \item{\textbf{Full Model.} Includes all 200 nodes in original XFOIL spacing and velocity data (Figure \ref{fig:TurbPlots}.A,B).}

    \item{\textbf{Alternate Node Configuration.} Original model but with the velocity matrix arranged as LE-to-TE top, LE-to-TE bottom rather than the TE-LE-TE wraparound ordering of the original.}

    \item{\textbf{Bottom Surface in Positive $\alpha$.} This model contained only the original 100 bottom surface nodes and only considered the positive angles from 0.5$\degree$ to 16$\degree$, a mostly laminar flow.} 

    \item{\textbf{Top Surface in Negative $\alpha$.} This model contained only the original 100 top surface nodes and only considered the negative angles from -10$\degree$ to 0.0$\degree$.}

    \item{\textbf{Full Surface NACA 0012.} This symmetric airfoil was evaluated to determine if the best possible and SSPOP DPs would be approximately symmetrical.}
\end{enumerate}

\subsubsection{Analysis and Summary of NACA 4415}

Table \ref{tab:Results4415_2} lists the DP sensor locations (as depicted in Figure \ref{fig:PhaseIplots3}), predictive performance by RMSE of $\alpha$, and ranking against all possible DPs for each model variant. With the exception of the NACA 0012 model (J), all DP node numbers are listed by their original full model numbering (Figure \ref{fig:NACA4415}). Models A through F are ranked in Table \ref{tab:Results4415_2} in order of their SSPOP DP ranking within all possible DPs for each model. Interestingly, the full model (F) was the worst performer among these, though the SSPOP+S for the full model still performed very well. This indicates that the SSPOP algorithm had greater success finding flow features related to transition when there are is a sharp discontinuity at the transition point. The constrained model with turbulent signals set to the average values (A) far outperformed all models in SSPOP DP ranking, indicating that some corrective overwrite of turbulent signals might be useful in FBF.

Model G is comprised of the same airflow velocity data as the full model (F) but ordered in a different sequence. This was done to test whether SSPOP behaves differently based on matrix structure given that the linear regression models are independent of the matrix structure. While this was confirmed (the best possible DP and RMSE was the same for both), the SSPOP and SSPOP+S results were slightly different for the two models. The difference does not appear to be significant.

Models H and J were created to evaluate the effectiveness of separate sensor placement optimization for the bottom and top surfaces, respectively. Since laminar flow dominates on the top surface for negative $\alpha$ and on the bottom surface for positive $\alpha$, a wing with a dual-FBF system could alternate between the use of top and bottom sensors as needed to get clean data. The linear regression models are much better over these smaller sign-homogeneous ranges of $\alpha$. The best possible performance for these two models far outperforms the rest, as expected. However, while the SSPOP for the bottom surface performed adequately by DP rank, the SSPOP for the top surface (27.9$\%$) was the only major outlier. This outlier remains unexplained after multiple checks of the SSPOP algorithm and data.

The full symmetric NACA 0012 model (J) was investigated to confirm the hypothesis from Phase I.a that sensors should be placed symmetrically across the chord line. Referring to Figure \ref{fig:PhaseIplots3}.J, it is evident that the best possible DP is symmetric and the SSPOP and SSPOP+S DPs are nearly symmetric, as expected. The implication is that a symmetric airfoil need only be analyzed on one surface only.

\begin{figure}
\centering 
\includegraphics[width=\textwidth]{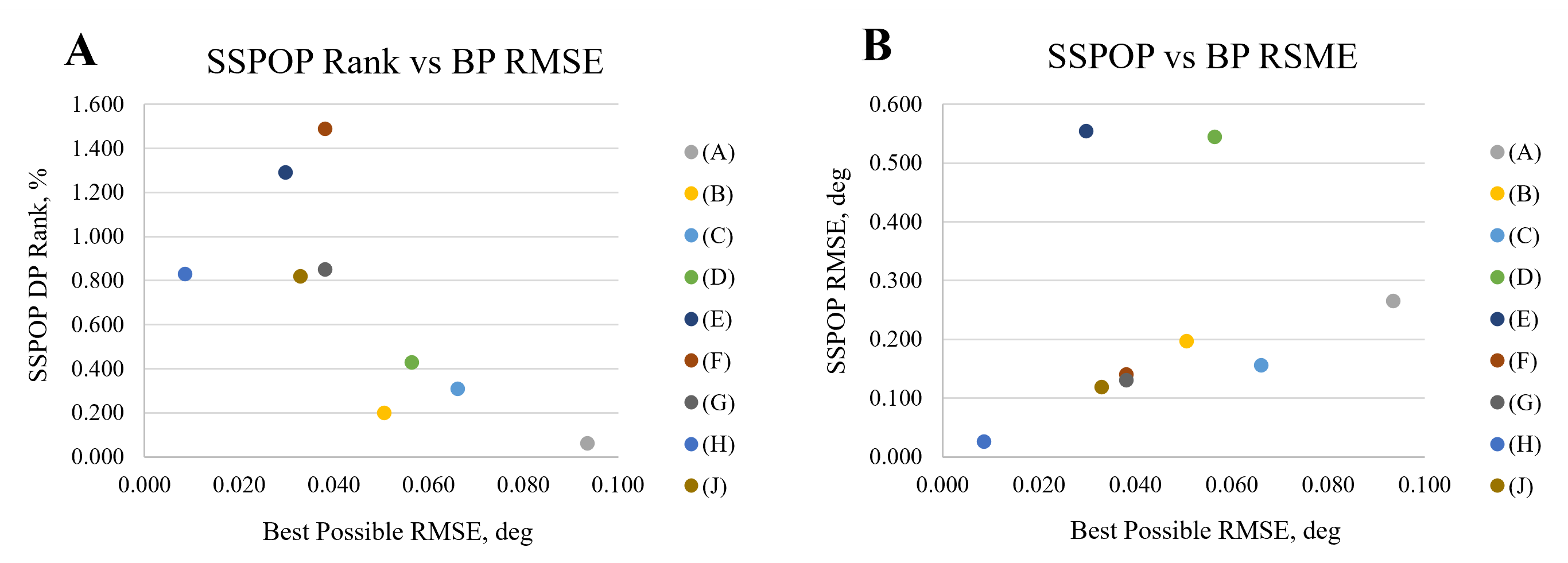}
\caption{Phase I absolute and relative results. \hspace{.2mm} A. SSPOP DP Ranking as percentage of all DPs versus the RMSE of the best possible DP. \hspace{.4mm} B. RMSE of the SSPOP DP versus the best possible DP. Outlier Model (I) is not included.}
\label{fig:rankings3}
\end{figure}

Figure \ref{fig:rankings3} shows the BP RMSE, the SSPOP RMSE, and the ratios between them for each model. There is a positive correlation with the BP performance and the SSPOP DP performance. This means that in cases where the BP DP cannot be found (e.g. when \textit{Q} and \textit{N} are large), models whose SSPOP DP performs best will also have the best solutions to be found by a search. The SSPOP ranking follows a exponentially decreasing trend with increasing best possible RMSE. This can mean that SSPOP works best on models which are designed around good linear regression performance. These SSPOP-to-BP RMSE ratios for models D and E (Turb. = 0) appear to be outliers on an otherwise linear trend. 

\subsubsection{Summary of Phase I}
The SSPOP algorithm was able to select a DP in or near the top 1$\%$ of all possible DPs as ranked by RMSE of $\alpha$ prediction. Then, using the SSPOP DP performance as a target, a search algorithm quickly finds a DP with an RMSE very near that of the best possible DP. Altogether, the SSPOP+S method takes approximately two minutes versus ten hours on a high-powered laptop for a combanitorial evaluation of the four-sensor case for a 200-node airfoil. Similar performance and cost savings can be expected for cases with much higher sensor numbers where the BP DP cannot be identified. The necessity for a data-reduction approach such as SSPOP becomes readily apparent when considering the combanitorial problem posed by a few dozen sensors and several thousand nodes on a 3D wing.

\subsection{Phase II - Wing Section} \label{Phase2}

\begin{figure}
\centering 
\includegraphics[width=\textwidth]{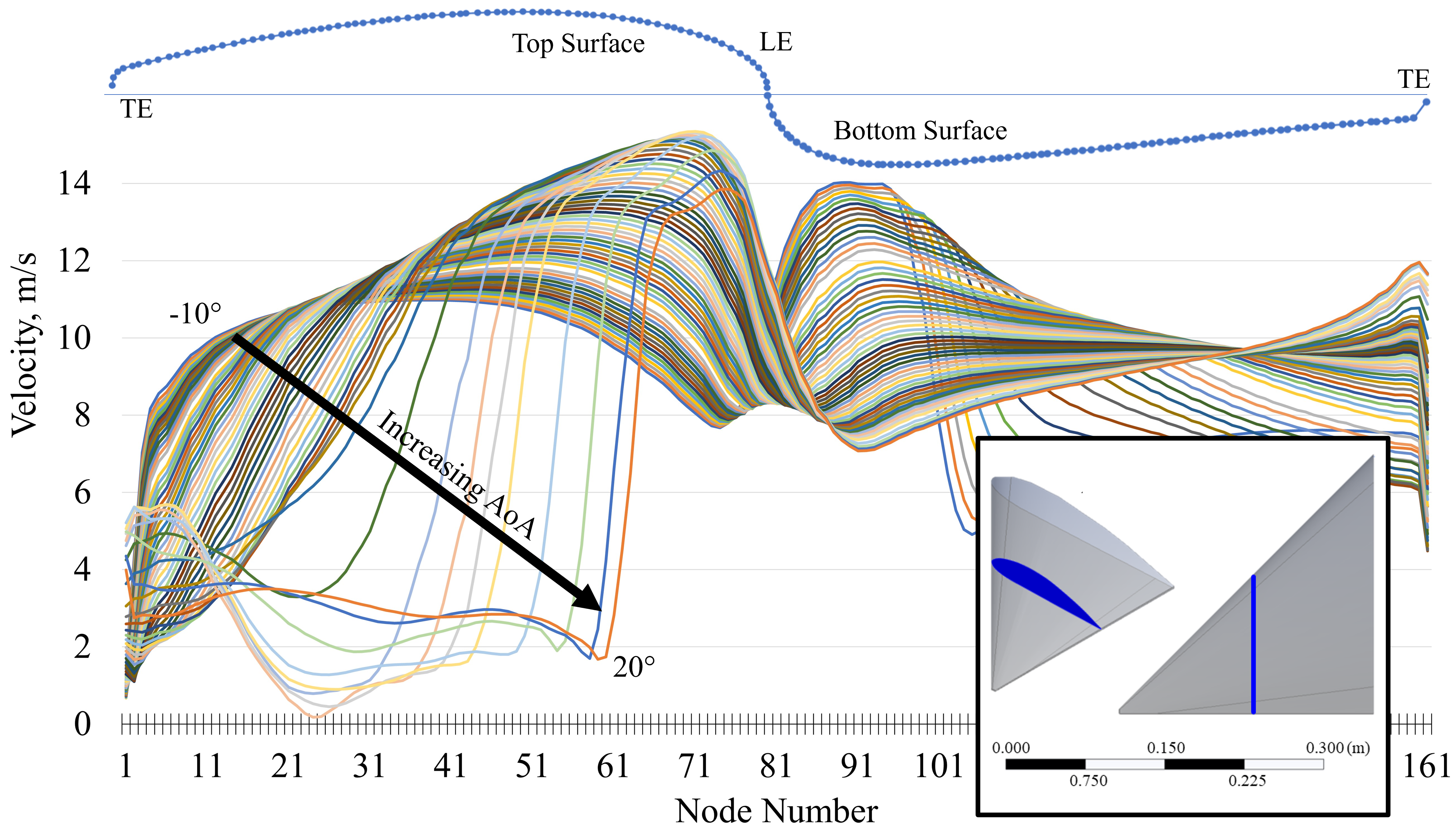}
\caption{Phase II delta wing slice velocity profiles showing notable vortex discontinuities. Node locations are depicted at the top. Compare with clean profiles from infinite wing slice from Phase I.b ( Figure \ref{fig:TurbPlots}.B). Inset: 45$\degree$ sweep NACA 4415 delta wing showing location of airfoil slice.}
\label{fig:PhaseIIprofiles3}
\end{figure}

Phase II entailed calculating 3D flow data with CFD for a delta wing (see Section \ref{CFD}) and performing SSPOP analysis for placement of AHSs along a 2D chordwise slice of the wing. The complexity of this flow is apparent when comparing the velocity profiles of this 3D wing slice (Figure \ref{fig:PhaseIIprofiles3}) against the clean profiles of the infinite wing slice (Figure \ref{fig:TurbPlots}.B). The SSPOP algorithm was applied to the data matrix of velocity magnitudes along this airfoil section in the same way as in Phase I. The full model for Phase II had 164 nodes and the constrained model, with 20 percent of the TE removed to represent a control surface, had 129 nodes. Airflow velocity magnitude data were computed for 56 angles of attack from -10$\degree$ to 20$\degree$, rather than up to only 16$\degree$ as in Phase I, in order to better capture discontinuous behavior due to separation and vortices close to stall. 

\begin{figure}
\centering 
\includegraphics[width=\textwidth]{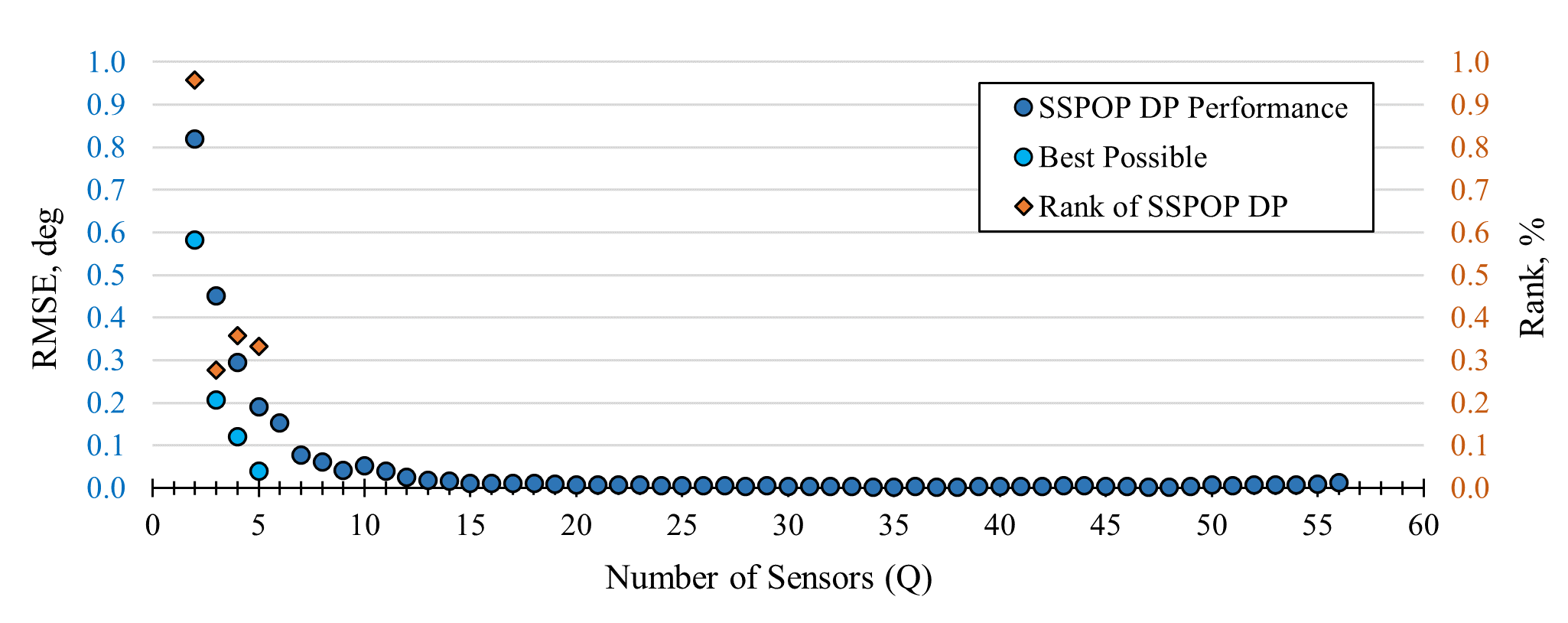}
\caption{Phase II Results (constrained model). SSPOP RMSE, BP RMSE, and SSPOP DP Ranking for 2 to 5 sensors, SSPOP RMSE for 12 to 56 sensors.}
\label{fig:PhaseIIResults4}
\end{figure}

\begin{figure}
\centering 
\includegraphics[width=\textwidth]{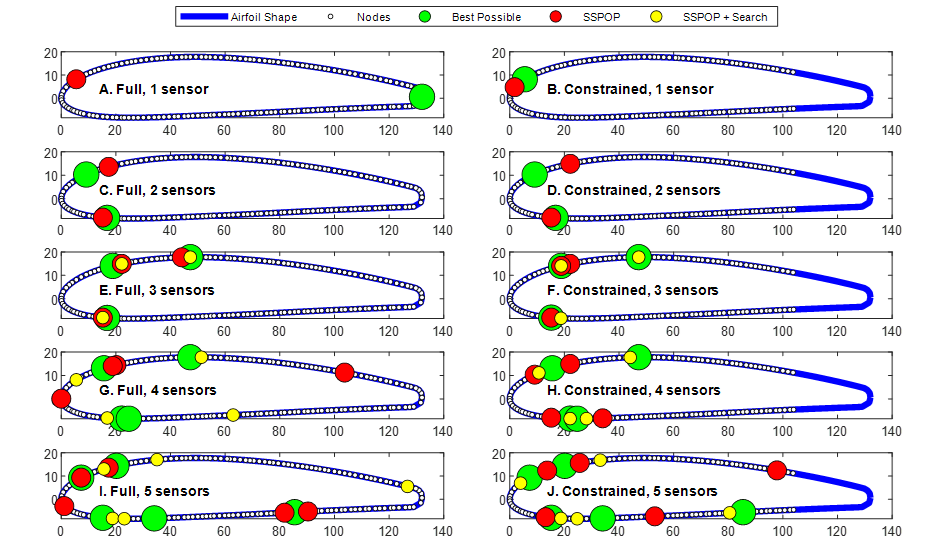}
\caption{Phase II DPs for 1 through 5 sensors. Colored dots show location of each sensor for each DP. Nodes or candidate sensor locations are represented by white dots.}
\label{fig:PhaseIIplots2}
\end{figure}

The results of $\alpha$ prediction by RMSE for Phase II constrained model are shown in figure \ref{fig:PhaseIIResults4}. The Best Possible and SSPOP DP performances and the ranking of the SSPOP DP for two through five sensors, while the SSPOP DP performance is shown for every Q through 56 sensors. The full model results followed similar trends. The DP sensor locations are plotted in Figure \ref{fig:PhaseIIplots2}. The SSPOP DP for the full model with 4 sensors was a surprising outlier; its rank of 2.519$\%$ was unexpectedly high while the rankings for the constrained model with 4 sensors and both models with 5 sensors were as expected. Interestingly, the SSPOP+S process improved the outlier performance by one order of magnitude, as it does for the others in Phase II.

An analysis of SSPOP-only performance was carried out for 6 through 56 sensors. Figure \ref{fig:PhaseIIResults4} shows the SSPOP RMSE decreasing exponentially up to about 15 sensors. The minimum occurs around 38 sensors, which is well above the realistic maximum \textit{Q} for an airfoil of this size. The expected increase in RMSE near 56 sensors is observed as the linear regression models become overfit. The full model DPs generally outperformed those of the constrained model, but the difference is minimal because both SSPOP and the BP DPs tend to be on nodes forward of mid-chord where there were no constraints: Figure \ref{fig:PhaseIIplots2} shows that the Best Possible DP for both models is exactly the same (except in the trivial 1-sensor case where the DP was at the TE), and the SSPOP and SSPOP+S DPs are nearly the same. As with Phase I, it can be concluded that the optimal placement for AHSs will be near, but not on, the LE and symmetric over the chord line. 

\section{Discussion and Future Work} \label{diss}

The SSPOP algorithm is capable of quickly finding a DP of AHS placements which predicts $\alpha$ with linear regression to well within 0.10 degrees accuracy over a wide range of $\alpha$ for 2D airfoils. The wide scope of this algorithm was demonstrated by evaluating several airfoil model configurations in 2D and 3D flows and with ideal and artificially noisy data. The performance of these SSPOP DPs ranks within the top 1$\%$ of all possible DPs for those cases where a brute force analysis was performed, granting confidence that similar performance can be expected for any arbitrary number of sensors and nodes. Therefore, when a brute force search is infeasible and the true optimum DP cannot be found (i.e. models with large \textit{N} and/or \textit{Q}), we can use the SSPOP DP RMSE as a top-one-percent starting point. A search can then be performed to quickly find a DP which performs an order of magnitude better.

The real-world success of the SSPOP algorithm depends on the quality of the data input. In this work we relied on computed flow velocity magnitudes and, in some cases, made assumptions to modify the data according to expected AHS sensory capabilities. Future iterations of the SSPOP algorithm can be improved by using wind tunnel tests to characterize the performance of individual and arrayed AHSs in various flow conditions. The physical AHS described in this paper which inspired this algorithm is currently being experimentally characterized in preparation for future SSPOP validation experiments \cite{Izquierdo2024ImprovingCalibration}.

This work primarily considered $\alpha$ as the objective for prediction, and it was the only flight condition variable. A broader analysis for a more robust FBF application should include variable freestream velocity and a range of sideslip angles as well as angles of attack. A practical FBF control system must be able to provide accurate flight state predictions over the entire range of expected conditions.  

\section{Summary Remarks}

This paper described the development of a novel approach for identifying the location of a sparse set of velocity magnitude sensors for predicting aerodynamic parameters. The models included NACA 0012 and 4415 airfoils and a canonical NACA 4415 45$\degree$ swept blunt-edged delta wing. The sparse sensor placement optimization for prediction (SSPOP) algorithm was introduced and used to find a near-optimal location of a handful of sensors for predicting $\alpha$ (and in one case \(c_l\), \(c_d\), and \(c_m\)) by a sparse linear regression model. A brute force search of all possible DPs, where possible, enabled relative evaluation of SSPOP DP performance. Additionally, a sensitivity analysis of node numbers and regression ratios confirmed the hypothesis that a greater set of candidate sensor positions enabled more accurate predictions by a given number of sensors. The SSPOP algorithm was also tested with artificial noise and was able to extract useful features for sensor placement in noisy conditions. 

In all cases for 2 or more sensors, the SSPOP algorithm performed very well for predicting $\alpha$, always finding a DP within the top 1$\%$ of all possible DPs by RMSE. Adding a search procedure after SSPOP yielded a DP within the top 0.01$\%$ in a few minutes versus the several hours required for a brute force search. The SSPOP algorithm is highly adaptable. It can be applied in two or three dimensions for any shape and node configuration and for pressure, strain, or any other FBF-relevant sensors. By quickly and easily guiding the placement of sensors, the SSPOP algorithm can support the implementation of FBF control in the design and operation of next-generation aircraft.

\section*{Reproducible Research} \label{repro}
A Matlab code supplement\footnote{Available on GitHub at URL \hyperlink{https://github.com/AlexHbeck/SSPOP/tree/ef9f7d685b0741b194eac69904ad892f60a4878d}{https://github.com/AlexHbeck/SSPOP/tree/ef9f7d685b0741b194eac69904ad892f60a4878d}} is available for reproducing some of the results in this manuscript, including the SSPOP algorithm code, flow datasets for Phase II, Matlab scripts to recreate the design point airfoil plots, and detailed flow charts of the SSPOP algorithm. 

\section*{Acknowledgments}
The views and conclusions contained herein are those of the authors and do not necessarily represent official policies or endorsements, either expressed or implied, of the U.S. Air Force, the U.S. Department of Defense, or the U.S. Government.  This paper is cleared for public release, case number 88ABW-2023-0966. This material is declared a work of the U.S. Government and is not subject to copyright protection in the United States. See also AIAA Rights and Permissions \hyperlink{https://www.aiaa.org/publications/Publish-with-AIAA/Rights-and-Permissions}{www.aiaa.org/randp}.

\bibliography{references}

\end{document}